\documentclass[aps,preprint,showpacs,amssymb,eqsecnum]{revtex4}
\usepackage{graphicx}% Include figure files
\begin{document}
\title
{Weak noise approach to the logistic map}
\author{Hans C. Fogedby}
\email{fogedby@phys.au.dk} \affiliation {
Department of Physics and Astronomy\\
University of Aarhus\\
Ny Munkegade, DK-8000 Aarhus C, Denmark\\
and\\
NORDITA, Blegdamsvej 17, DK-2100, Copenhagen {\O}, Denmark
}
\author{Mogens H. Jensen}
\email{mhjensen@nbi.dk}
\affiliation
{
Niels Bohr Institute for Astronomy, Physics, and Geophysics\\
Blegdamsvej 17, DK-2100, Copenhagen {\O}, Denmark
}
%\date{\today}
%\footnote{permanent address}
\begin{abstract}
Using a nonperturbative weak noise approach we investigate the
interference of noise and chaos in simple 1D maps. We replace the
noise-driven 1D map by an area-preserving 2D map modelling the
Poincare sections of a conserved dynamical system with unbounded
energy manifolds. We analyze the properties of the 2D map and draw
conclusions concerning the interference of noise on the nonlinear
time evolution. We apply this technique to the standard
period-doubling sequence in the logistic map. From the 2D
area-preserving analogue we, in addition to the usual
period-doubling sequence, obtain a series of period doubled cycles
which are elliptic in nature. These cycles are spinning off the
real axis at parameters values corresponding to the standard
period doubling events.
\end{abstract}
\maketitle

%%%%%%%%%%%%%%%%%%%%%%%%%%%%%%%%%%%%%%%%%%%%%%
%%%%%%%%%%%%%%%%%%%%%%%%%%%%%%%%%%%%%%%%%%%%%%
\section{\label{secintro}Introduction}
%%%%%%%%%%%%%%%%%%%%%%%%%%%%%%%%%%%%%%%%%%%%%%
%%%%%%%%%%%%%%%%%%%%%%%%%%%%%%%%%%%%%%%%%%%%%%
The noise-induced escape from an equilibrium state is a
fundamental problem in many areas in science. The classical work
goes back to Kramers \cite{Kramers40} who computed the transition
rate from a single potential well; see also the review in Ref.
\cite{Haenggi90}. More recently, the understanding and description
of nonlinear dynamical systems exhibiting e.g. period-doubling,
intermittency, and chaos, have led to a renewed interest in the
influence of noise on the behavior of such systems
\cite{Graham84,Graham84a,Graham85,Graham86,Graham89,Moss89}.

In this context the generation of chaotic motion in low
dimensional time-discrete systems
\cite{Collet80,Grossmann77,Feigenbaum78,Feigenbaum79} offers a
particularly simple case lending itself to the analysis of the
influence of noise. Thus a variety of phenomena have been
investigated such as the noise-induced shift, broadening, and
suppression of bifurcations \cite{Haken81,Linz86}, the scaling
properties of Lyapunov exponents \cite{Crutchfield81,Shraiman81}
and the invariant density \cite{Reimann91a,Graham91} near the
threshold for chaos, the scaling of intermittency \cite{Hirsch82},
and escape from locally stable states
\cite{Arecchi84,Grassberger89,Talkner87,Crutchfield82,
Reimann91a,Reimann91b,Reimann94,Reimann95}; see also recent work
on the application of periodic orbit theory to noisy maps
\cite{Cvitanovic98,Cvitanovic99a,Cvitanovic99b}.

For the purpose of studying the interference of stochastic noise
with the nonlinear behavior of dynamical systems, the simplest
case is the influence of noise on 1-D maps of the generic type
\begin{eqnarray}
x_{n+1}=f(x_n).
\label{map}
\end{eqnarray}
Here $n$ is the discrete time index and the map $f(x)$ defines the
nonlinear discrete time evolution. In the linear case $f(x)\propto
x$ the map is readily analyzed; it possesses an attractive or
repulsive fixed point  at $x=0$ depending on the slope, and does
not exhibit chaos. In the nonlinear case, where $f(x)$ possesses
one or several differentiable maxima, the map generates the
well-known period-doubling sequence to a chaotic state beyond a
critical value of the control parameter characterizing the map
\cite{Feigenbaum78,Feigenbaum79}. In the singular case of for
example the tent or the shift map, the period-doubling sequence is
absent and the time evolution becomes chaotic beyond a critical
control parameter value, see e.g. Refs.
\cite{Ott93,Schuster89,Strogatz94} .

Assuming that the deterministic map in Eq. (\ref{map}) describes
the dynamical evolution of a nonlinear system in an effectively
reduced phase space, the presence of stochastic degrees of freedom
originating from the environment can typically be modelled by an
additive noise term and we are led to the stochastic map
\begin{eqnarray}
x_{n+1}=f(x_n)+\xi_n; \label{nmap}
\end{eqnarray}
a discrete version of the usual Langevin equation for stochastic
processes \cite{vanKampen92}.

Here $\xi_n$ indicates the applied noise which we for simplicity
assume has a Gaussian distribution
\begin{eqnarray}
P(\xi_n)=\frac{1}{\sqrt{2\pi\Delta}}
\exp\left[-\frac{\xi_n^2}{2\Delta}\right],
\label{disnoise}
\end{eqnarray}
and is correlated according to
\begin{eqnarray}
\langle\xi_n\xi_m\rangle = \Delta\delta_{nm},
\label{corr}
\end{eqnarray}
where $\Delta$ is the noise strength and $\langle\cdots\rangle$
indicates an average over the noise ensemble.

The map $f$ acts like a nonlinear filter on the known white noise
input fluctuations $\xi_n$ and the fundamental issue becomes that
of determining the stochastic properties of the stochastic
variable $x_n$. The natural small parameter is the noise strength
$\Delta$ which enters in a nonperturbative manner as indicated by
the form of $P(\xi)$ in Eq. (\ref{disnoise}). Physically, the case
of vanishing noise $\Delta=0$, yielding the dissipative map in Eq.
(\ref{map}) for $\xi=0$, is quite distinct from the case of weak
noise $\Delta\sim 0$, where on a sufficiently long time scale
(labelled by the iteration index n) the noise $\xi_n$ drives the
variable $x_n$ into a stochastic state. The issue is to understand
how the stochastic noise interferes with the nonlinear behavior.

In recent work we have elaborated on a nonperturbative canonical
phase space approach based on the Freidlin-Wentzel formulation
\cite{Freidlin98} or, alternatively, a saddle point approximation
to the Martin-Siggia-Rose method in its functional form
\cite{Martin73,Baussch76} and have applied this scheme to the
stochastic Kardar-Parisi-Zhang equation \cite{Medina89,Kardar86}
describing the evolution of a growing interface
\cite{Fogedby95,Fogedby98b,Fogedby99a,Fogedby02a,Fogedby03b}.

It is characteristic of the canonical phase space formulation that
the stochastic evolution equation is replaced by coupled
Hamiltonian equations of motion, the noise being replaced by a
canonical momentum variable, and that the transition probability
is obtained from the action associated with a solution (an orbit
in phase space) of the Hamilton equations. The formulation is a
weak noise approximation in the same spirit as the well-known WKB
approximation in quantum mechanics and the noise strength enters
in the same way as the Planck constant in the WKB scheme
\cite{Landau59c,Reichl87}.

In the present paper we attempt to adapt the above singular weak
noise scheme to a stochastic map. We again find a characteristic
variable doubling in the sense that the stochastic map in Eq.
(\ref{nmap}) is replaced by the 2D map
\begin{eqnarray}
&&x_{n+1}=f(x_n)+p_n, \label{2d1}
\\
&&p_{n-1}= f'(x_n)p_n, \label{2d2}
\end{eqnarray}
where the noise $\xi_n$ in Eq. (\ref{nmap}) is replaced by the new
noise variable $p_n$. Note also that the equation for $p_n$
iterates `backwards' in time. The transition probability from
$x_n$ to $x_m$ in time $N$, $P(x_n\rightarrow x_m,N)$ is given by
\begin{eqnarray}
P(x_n\rightarrow
x_m,N)=\Omega(N)^{-1}\exp\left[-\frac{S(x_n\rightarrow x_m,N)}
{\Delta}\right], \label{trans}
\end{eqnarray}
with dynamic partition function
\begin{eqnarray}
\Omega(N)=\sum_{x_m}\exp\left[-\frac{S(x_n\rightarrow x_m,N)}
{\Delta}\right]. \label{dynpar}
\end{eqnarray}
The action $S$ has the form
\begin{eqnarray}
S=\frac{1}{2}\sum_{n=1}^Np_n^2.
\label{action}
\end{eqnarray}
The formal scheme is thus straightforward. From the 2D map given
by Eqs. (\ref{2d1}) and (\ref{2d2}) we extract an orbit from $x_n$
to $x_m$ traversed in $N$ steps. The initial and final condition
on $x$ together with  the time span $N$ then defines $p_n$ and $P$
follows from Eq. (\ref{trans}) by means of the action $S$
evaluated in Eq. (\ref{action}).
%%%%%%%%%%%%%%%%%%%%%%%%%%%%%%%%%%%%%%%%%%%%%%
%%%%%%%%%%%%%%%%%%%%%%%%%%%%%%%%%%%%%%%%%%%%%%
\section{\label{secweak}Weak noise scheme}
%%%%%%%%%%%%%%%%%%%%%%%%%%%%%%%%%%%%%%%%%%%%%%
%%%%%%%%%%%%%%%%%%%%%%%%%%%%%%%%%%%%%%%%%%%%%%
\subsection{\label{gen} Generalities}
%%%%%%%%%%%%%%%%%%%%%%%%%%%%%%%%%%%%%%%%%%%%%%
%%%%%%%%%%%%%%%%%%%%%%%%%%%%%%%%%%%%%%%%%%%%%%
Generally, the stochastic properties of the noisy map in Eq.
(\ref{nmap}) can be extracted from the generator
\begin{eqnarray}
Z(\{\mu_n\})=\left\langle\int\prod_n
dx_ne^{i\sum_n\mu_nx_n}\right\rangle, \label{gen1}
\end{eqnarray}
where $x_n$ is driven by the map and $\langle\cdots\rangle$
indicates an average over the implicit noise dependence.
Implementing the map by means of a delta function constraint which
is subsequently exponentiated, and averaging over the noise
according to Eq. (\ref{disnoise}) we obtain
\begin{eqnarray}
Z(\{\mu_n\})  &&\propto \int\prod_nd\xi_ndx_ne^{i\mu_nx_n}
\delta(x_{n+1}-f(x_n)-\xi_n)e^{-\xi_n^2/2\Delta} \nonumber
\\
&&\propto \int\prod_nd\xi_ndx_ndp_ne^{i\mu_nx_n}
e^{ip_n(x_{n+1}-f(x_n)-\xi_n)}e^{-\xi_n^2/2\Delta}.
\label{gen2}
\end{eqnarray}
Finally, integrating over the noise variable $\xi_n$ and making
the replacement $i\Delta p_n\rightarrow -p_n$ we arrive at the form
\begin{eqnarray}
Z(\{\mu_n\})\propto\int\prod_n dx_ndp_n e^{i\mu_n x_n}
e^{-S/\Delta}, \label{gen3}
\end{eqnarray}
where the action $S$ is given by
\begin{eqnarray}
S=\sum_n p_n\left(x_{n+1} - f(x_n)-\frac{1}{2}p_n\right).
\label{action2}
\end{eqnarray}
In the weak noise limit $\Delta\rightarrow 0$ the dominant part of
the functional integral in Eq. (\ref{gen3}) is determined by the
orbits minimizing the action $S$ and we obtain a principle of
least action yielding the condition $\delta S=0$ subject to
variations of $x_n$ and $p_n$. Setting $\delta S/\delta
x_n=p_{n-1}- p_nf'(x_n)=0$ and $\delta S/\delta
p_n=x_{n+1}-f(x_n)-p_n=0$ we then obtain the equations of motion
in the form of the 2D map in Eqs. (\ref{2d1}) and (\ref{2d2}). For
the optimal orbit traversed in time $N$ we also obtain by
inserting Eq. (\ref{2d1}) in Eq. (\ref{action2}) the action in Eq.
(\ref{action}). From $Z(\{\mu_n\})$ we extract the transition
probability from $x_n$ to $x_m$ in time $N$ given by Eq.
(\ref{trans}).

In order to evaluate the transition probability from $x_n$ to
$x_m$ in time $N$ we then have to first solve the 2D map given by
Eqs. (\ref{2d1}) and (\ref{2d2}) subject to the conditions
$x_q=x_n$ for $q=1$ and $x_q=x_m$ for $q=N$, $p_q$ being a slaved
variable; secondly, evaluate the action $S$ for this specific
orbit, and finally evaluate $P$ according to Eq. (\ref{trans}). We
note that the delta function constraint implementing the map gives
rise to the additional noise variable $p_n$. We also remark that
the equation for $p_n$ iterates backwards in time. Solving for
$p_n$ we can express the 2D map in a form iterating forward in
time
\begin{eqnarray}
&&x_{n+1}=f(x_n)+p_n, \label{eq11}
\\
&&p_{n+1}=\frac{1}{f'(x_{n+1})}p_n. \label{eq22}
\end{eqnarray}
Eliminating $p_n$ from Eqs. (\ref{eq11}), (\ref{eq22}), and
inserting in  Eq. (\ref{action}) we can also express $S$ in the
form
\begin{eqnarray}
S=\frac{1}{2}\sum_{n=1}^{N}[x_{n+1}-f(x_n)]^2,
\end{eqnarray}
in combination with the two-step recursion formula
\begin{eqnarray}
x_n-f(x_{n-1})-(x_{n+1}-f(x_n))f'(x_n)=0,
\end{eqnarray}
thus making contact with the work in Refs.
\cite{Haken81,Linz86,Reimann91a}.
%%%%%%%%%%%%%%%%%%%%%%%%%%%%
\subsection{\label{con} Continuum limit - Symplectic structure}
%%%%%%%%%%%%%%%%%%%%%%%%%%%%
It is instructive to perform the continuum limit $x_n\rightarrow
x(t)$, $p_n\rightarrow p(t)$. Setting $x_{n+1}\approx x+dx/dt$ and
$p_{n-1}\approx p-dp/dt$ we obtain from Eqs. (\ref{2d1}) and
(\ref{2d2}) the coupled differential equation, the 2D flow
\begin{eqnarray}
&&\frac{dx}{dt} = -x + f(x) + p,
\label{eq13}
\\
&&\frac{dp}{dt} = p - pf'(x),
\label{eq23}
\end{eqnarray}
originating from the Hamiltonian
\begin{eqnarray}
H = p(-x + f(x)) + \frac{1}{2}p^2; \label{ham}
\end{eqnarray}
we note that compared with ordinary mechanics the Hamiltonian has
a momentum dependent potential and unbounded energy surfaces in
phase space. Correspondingly, the action in Eq. (\ref{action})
takes the symplectic form
\begin{eqnarray}
S = \int dt\left[p\frac{dx}{dt} - H\right].
\label{action4}
\end{eqnarray}
We conclude that in the continuum limit the weak noise approach to
the stochastic flow or Langevin equation
\begin{eqnarray}
\frac{dx}{dt} = -x + f(x) +\xi,
\label{lan}
\end{eqnarray}
with noise correlations
\begin{eqnarray}
\langle\xi(t)\xi(t')\rangle(t)=\Delta\delta(t-t'), \label{corr2}
\end{eqnarray}
yields a canonical structure described by the conserved 2D flow in
Eqs. (\ref{eq13}) and (\ref{eq23}). The phase space plot is
spanned by $x$ and $p$ and the orbits lie on the energy surfaces
given by Eq. (\ref{ham}). The zero-energy surfaces are spanned by
$p=0$, the transient manifold, and $p=2x-2f(x)$, the stationary
manifold. The transition probability from $x_1$ to $x_2$ in time
$T$ is given by $P\propto\exp[-S/\Delta]$ and is evaluated by
solving the equations of motion (\ref{eq13}) and (\ref{eq23}) for
an orbit from $x_1$ to $x_2$ traversed in time $T$ and computing
the action $S$ in Eq. (\ref{action4}). At long times the orbit
migrates to the zero-energy manifolds and pass by the saddle point
at the intersection of the two manifolds yielding an ergodic
stationary state. Expressing the Langevin equation in the form
\begin{eqnarray}
\frac{dx}{dt} = -\frac{1}{2}\frac{dF}{dx} +\xi,
\label{lan2}
\end{eqnarray}
where the free energy is given by
\begin{eqnarray}
F=x^2-2\int^xf(y)dy,
\label{free}
\end{eqnarray}
it follows that the system can attain a stationary state with
probability distribution
\begin{eqnarray}
P_{\text{st}} \propto\exp\left[-\frac{F}{\Delta}\right].
\label{stat}
\end{eqnarray}
This result also follows from $S=\int dt p dx/dt$ on the
zero-energy manifold setting $p=2x-2f(x)$. We shall not pursue
this analysis further here but refer to Refs.
\cite{Fogedby95,Fogedby98b,Fogedby99a,Fogedby02a,Fogedby03b} for
more details. The main point is that in  the continuum limit the
weak noise approach applied to a noisy 1D flow yields a symplectic
structure with conserved 2D flow. On the other hand, the 2D flow
does not exhibit chaotic behavior which is the main issue under investigation
here.
%%%%%%%%%%%%%%%%%%%%%%%%%%%%
\subsection{General discussion of 2D map}
%%%%%%%%%%%%%%%%%%%%%%%%%%%%
The forward iterating 2D map in Eqs. (\ref{eq11}) and (\ref{eq22})
is not area-preserving. This follows from the Jacobian
$J=\partial(x_{n+1},p_{n+1})/\partial(x_n,p_n)=f'(x_n)/f'(x_{n+1})$
which only in the continuum limit approaches unity, implying
area-preservation as discussed in Sec. \ref{con}. Only in the
vicinity of the fixed points where the orbits slow down does the
continuum limit apply and we obtain the conservation of area.
However, introducing the new noise variable $\pi_{n+1}=p_n$
shifted by one itration and coinciding with $p_n$ in the
continuum limit we obtain the 2D map
\begin{eqnarray}
&&x_{n+1}=f(x_n)+\frac{1}{f'(x_n)}\pi_n,
\\
&&\pi_{n+1}=\frac{1}{f'(x_n)}\pi_n,
\end{eqnarray}
which has Jacobian $J=1$ and is thus area-preserving.
%%%%%%%%%%%%%%%%%%%%%%%%%%%%%%%%%%%%%%%%%%%%%%
%%%%%%%%%%%%%%%%%%%%%%%%%%%%%%%%%%%%%%%%%%%%%%%%
\subsubsection{Fixed Point Structure}
%%%%%%%%%%%%%%%%%%%%%%%%%%%%%%%%%%%%%%%%%%%%
%%%%%%%%%%%%%%%%%%%%%%%%%%%%%%%%%%%%%%%%%%%
The fixed point structure of the 2D map follows from the equations
\begin{eqnarray}
&&x^\ast = f(x^\ast) + \frac{\pi^\ast}{f'(x^\ast)},
\\
&&\pi^\ast =\frac{\pi^\ast}{f'(x^\ast)},
\end{eqnarray}
with solutions
\begin{eqnarray}
&&(x^\ast,\pi^\ast)=(x_1^\ast,0),
\\
&&(x^\ast,\pi^\ast)=(x_2^\ast,\pi^\ast_2)=(x_2^\ast,x_2^\ast-f(x_2^\ast),
\end{eqnarray}
where $x_1^\ast$ is a fixed point of the deterministic 1D map and
$x_2^\ast$ is a solution of $f'(x_2^\ast)=1$ (for simplicity we
are assuming only one solution), i.e.,
\begin{eqnarray}
&&x_1^\ast=f(x_1^\ast),
\\
&&f'(x_2^\ast)=1.
\end{eqnarray}
The 2D conserved map is thus characterized by at least two fixed
points whose stability we proceed to analyze.

Expanding about a fixed point $(x^\ast,\pi^\ast)$ by setting
$x_n=x^\ast+\delta x_n$ and $\pi_n=\pi^\ast+\delta\pi_n$ we arrive
at the tangent map \cite{Reichl87}
\begin{eqnarray}
\left(
\begin{array}{c}
\delta x_{n+1}\\
\delta\pi_{n+1}
\end{array}
\right)=M\left(
\begin{array}{c}
\delta x_{n}\\
\delta\pi_{n}
\end{array}
\right), \label{tmap}
\end{eqnarray}
with stability matrix
\begin{eqnarray}
M=\left(
\begin{array}{cc}
f'(x^\ast)- \pi^\ast\frac{f''(x^\ast)} {f'(x^\ast)^2}
& \frac{1} {f'(x^\ast)}\\
-\pi^\ast\frac{f''(x^\ast)}{f'(x^\ast)^2}&\frac{1}{f'(x^\ast)}
\end{array}
\right). \label{smat}
\end{eqnarray}
Owing to area-preservation the determinant $|M|=1$ and the
properties of the fixed points are determined by the trace
\begin{eqnarray}
t=f'(x^\ast)+\frac{1}{f'(x^\ast)}-
\pi^\ast\frac{f''(x^\ast)}{f'(x\ast)^2}, \label{trace}
\end{eqnarray}
and the eigenvalues of $M$ are then given by
\begin{eqnarray}
\lambda_\pm=\frac{t}{2}\pm\sqrt{\left(\frac{t}{2}\right)^2-1}.
\label{eival}
\end{eqnarray}
The eigenvalues come in reciprocal pairs, $\lambda_+=
\lambda_-^{-1}$. For $|t|<2$  the eigenvalues form complex
conjugate pairs on the unit circle and the fixed point is
elliptic, for $t>2$ the fixed point is hyperbolic, and for $t<-2$
the fixed point is inversion hyperbolic. In the limiting case
$|t|=2$ the eigenvalues are degenerate, corresponding to the
parabolic case.

In the case of the {\em deterministic} fixed point
$(x^\ast,\pi^\ast)=(x_1^\ast,0)$, $x_1^\ast=f(x_1^\ast)$ we have
the trace, eigenvalues, and invariant manifolds
\begin{eqnarray}
&&t=f'(x_1^\ast)+\frac{1}{f'(x_1^\ast)},\\
&&\lambda_\pm=f'(x_1^\ast)^{\pm 1}, \\
&&(\delta x,\delta\pi)\propto (1,0),\\
&&(\delta x,\delta\pi)\propto(1,1-f'(x_1^\ast)^2).
\end{eqnarray}
Since $|t|\geq 2$ the fixed point is hyperbolic or inversion
hyperbolic; the degenerate case $|t|=2$ for $|f'(x_1^\ast)|=1$
corresponds to the parabolic case. In Fig.~{\ref{fig1} we have
depicted the orbits in the vicinity of the fixed points in the two
cases $|f'|>1$ and $|f'|<1$. We note that the orbits along the
invariant deterministic manifold $(1,0)$ corresponds to the 1D
map. In Fig.~\ref{fig2} we have shown the trace $t$ as a function
of the slope $f'(x_1^\ast)$ of the 1D map.
\begin{figure}
\includegraphics[width=0.9\hsize]
%{c:/user/manus/figs/162-217.eps}
{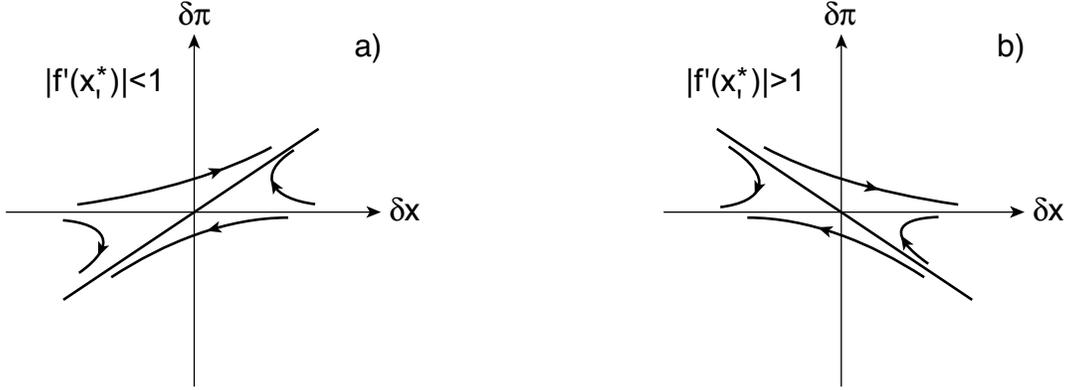} \caption{ We depict the orbits in the vicinity of
the deterministic hyperbolic fixed point ($x_1^\ast$,0). In a) we
show the orbits in the case $|f'(x_1^\ast)|<1$, where the fixed
point for the 1D map is stable. In b) we show the orbits for
$|f'(x_1^\ast)|>1$; the fixed point for the 1D map is unstable.}
\label{fig1}
\end{figure}
\begin{figure}
\includegraphics[width=0.9\hsize]
%{c:/user/manus/figs/162-220.eps}
{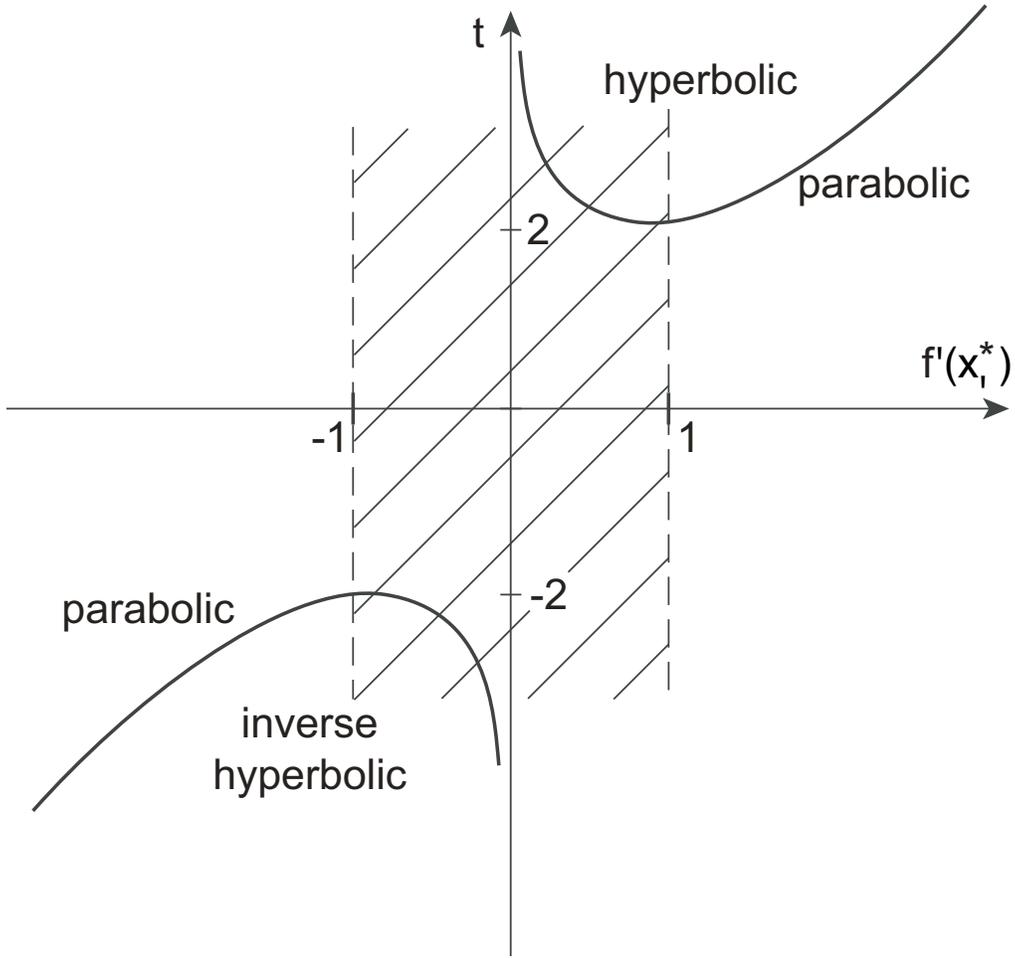} \caption{We plot the trace $t$ of the stability
matrix $M$ as a function of the slope $f'(x_1^\ast)$ for the
deterministic fixed point $(x_1^\ast,0)$. The shaded area
corresponds to the stability region of the 1D map.} \label{fig2}
\end{figure}

For the {\em noisy} fixed point
$(x^\ast,\pi^\ast)=(x_2^\ast,\pi_2)$,
$\pi_2=x_2^\ast-f(x_2^\ast)$, $f'(x_2^\ast)=1$, we obtain, setting
$f''_2=f''_2(x_2^\ast)$, trace, eigenvalues, and manifolds
\begin{eqnarray}
t&=&2-\pi_2^\ast f_2'',
\\
\lambda_\pm&=&\frac{\pi_2^\ast f_2''}{2}\pm \sqrt{
\left(\frac{\pi_2^\ast f_2''}{2}\right)^2-\pi_2^\ast f_2''},
\label{eigen}
\\
(\delta x,\delta\pi)&\propto& \left(1,\frac{\pi_2^\ast f_2''}{2} +
\sqrt{\left(\frac{\pi_2^\ast f_2''}{2}\right)^2+\pi_2^\ast
f_2''}\right),~~~~~~~~
\\
(\delta x,\delta\pi)&\propto& \left(1,\frac{\pi_2^\ast f_2''}{2} -
\sqrt{\left(\frac{\pi_2^\ast f_2''}{2}\right)^2+\pi_2^\ast
f_2''}\right),~~~~~~~~
\end{eqnarray}
and the character of the fixed point depends on the specific map
$f(x)$. In Fig.~{\ref{fig3} we have depicted the orbits in the
vicinity of the noisy fixed point in the two cases of a hyperbolic
and elliptic fixed point. In Fig.~{\ref{fig4} we have shown the
trace $t$ as a function of $\pi_2^\ast f''(x_2^\ast)$.
\begin{figure}
\includegraphics[width=0.9\hsize]
%{c:/user/manus/figs/162-221.eps}
{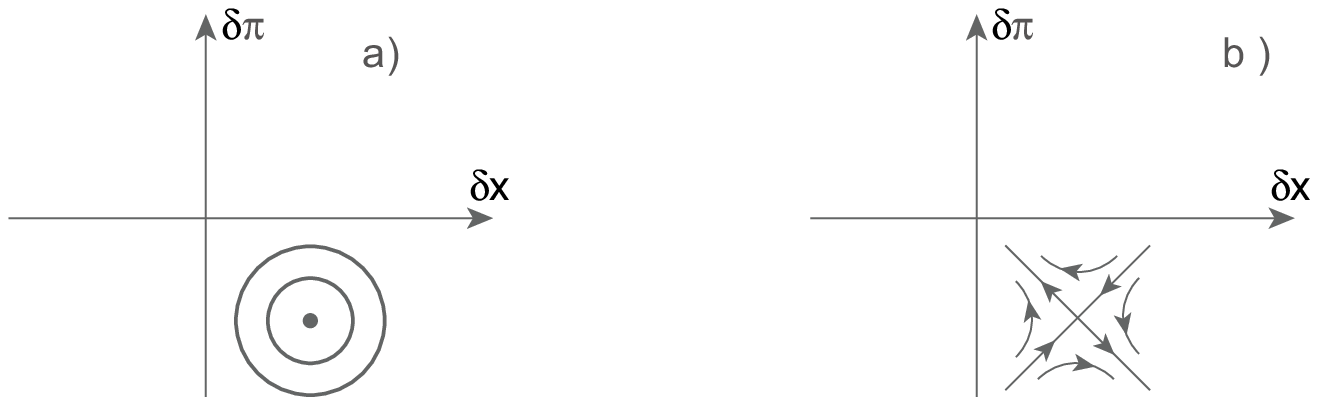} \caption{ We show the orbits in the vicinity of the
noisy fixed point. In a) we depict the elliptic case for
$|\pi_2^\ast f''(x_2^\ast)-2|<2$. In b) we show the hyperbolic
case for $\pi_2^\ast f''(x_2^\ast)>0$.} \label{fig3}
\end{figure}
\begin{figure}
\includegraphics[width=0.9\hsize]
{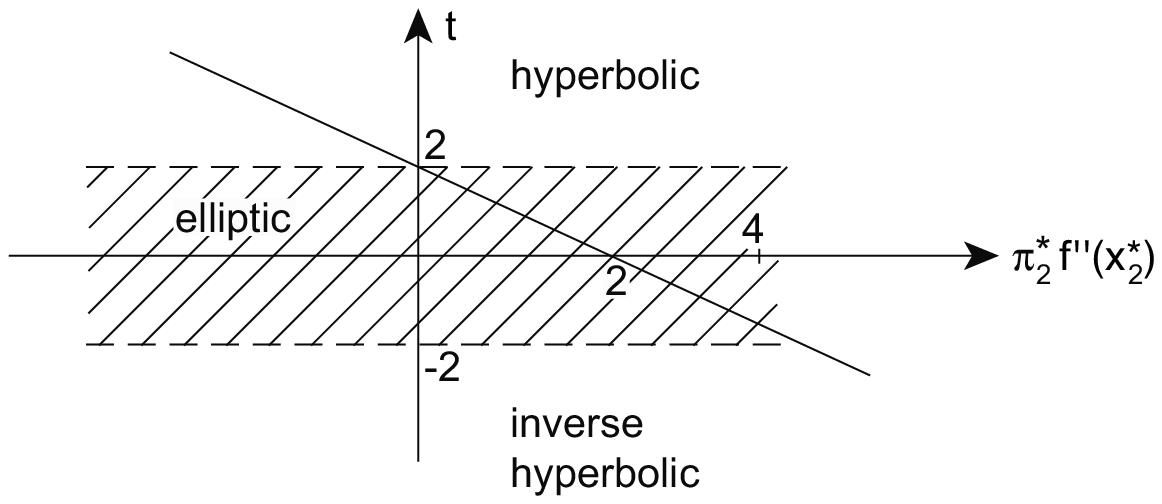}
%{c:/user/manus/figs/162-222.eps}
\caption{We plot the trace $t$ of the stability matrix $M$ as a
function $\pi_2^\ast f''(x_2^\ast)$. For $t>2$ the fixed point is
hyperbolic, for $t<-2$ the fixed point in inverse hyperbolic. For
$|t|<2$ the fixed point in elliptic.} \label{fig4}
\end{figure}
%
%%%%%%%%%%%%%%%%%%%%%%%%%%%%%%%%%%%%%%%%%%%%%%
%%%%%%%%%%%%%%%%%%%%%%%%%%%%%%%%%%%%%%%%%%%%%%%%
\subsection{Transition probabilities in
the vicinity of the fixed points}
%%%%%%%%%%%%%%%%%%%%%%%%%%%%%%%%%%%%%%%%%%%%
%%%%%%%%%%%%%%%%%%%%%%%%%%%%%%%%%%%%%%%%%%%
Leaving aside for a moment the issue of the nonlinear breakdown of
the orbit structure near the fixed point the weak noise scheme can
be applied in order to determine the transition probabilities.
%%%%%%%%%%%%%%%%%%%%%%%%%%%%%%%%%%%%%%%%%%%%%%
%%%%%%%%%%%%%%%%%%%%%%%%%%%%%%%%%%%%%%%%%%%%%%%%
\subsubsection{The deterministic fixed point}
%%%%%%%%%%%%%%%%%%%%%%%%%%%%%%%%%%%%%%%%%%%%
%%%%%%%%%%%%%%%%%%%%%%%%%%%%%%%%%%%%%%%%%%%
In the case of the deterministic fixed point
$(x^\ast,\pi^\ast)=(x_1^\ast,0)$, setting $a=f'(x_1^\ast)$ and
inserting the initial and final values $\delta x_1$ and $\delta
x_N$ and the time span $N$, the solution of the tangent map in Eq.
(\ref{tmap}) is
\begin{eqnarray}
&&\delta x_n=\frac {\delta x_N(a^{n-1}-a^{-n+1}) + \delta
x_1(a^{N-n}-a^{-N+n})}{a^{N-1}-a^{-N+1} }, \label{solx} \nonumber
\\
&& \\
 &&\delta \pi_n=(a^2-1)\frac{\delta x_N a^{1-n}-\delta x_1
a^{N-n}} {a^{N-1}-a^{-N+1}},\label{solp}
\end{eqnarray}
describing an orbit from $\delta x_1$ to $\delta x_N$ in time $N$
with $\delta\pi_n$ as a slaved variable.

For $|a|=|f'(x_1)|<1$ the fixed point is stable on the transient
invariant manifold $\delta \pi_n=0$. In the absence of noise the
motion on the manifold corresponds to damping, i.e., the 1D map is
line contracting. In the presence of noise for $\delta \pi_n\neq
0$ the orbits escape from the fixed point and approaches the
stationary invariant manifold $\delta\pi_n=(1-a^2)\delta x_n$;
this is the scenario depicted in Fig. 1 a). In the continuum limit
this behavior corresponds to the flow for the noise-driven
overdamped oscillator discussed in Ref. \cite{Fogedby99a}.

For $|a|>1$ the fixed point is unstable on the invariant manifold
$\delta\pi_n=0$. However, we note that in this case
$\delta\pi_n\rightarrow 0$. The fixed point is attractive along
the invariant noisy manifold $\delta\pi_n=-(a^2-1)\delta x_n$; the
orbit structure is depicted in Fig. 1 b). In the vicinity of the
hyperbolic fixed point the orbit slows down and an orbit from
$\delta x_1$ to $\delta x_N$ in time $N$ must pass asymptotically
through the fixed point for $N\rightarrow \infty$.

In order to evaluate the transition probability from  $\delta x_1$
to $\delta x_N$ in time $N$ we insert $\delta\pi_n=\delta p_{n-1}$
in the expression for the action in Eq. (\ref{action}). Performing
the sum yields the action
\begin{eqnarray}
S=\frac{1}{2}(a^2-1)(1-a^{-2N})\left(\frac{\delta x_N-\delta
x_1a^{N-1}}{a^{N-1}-a^{-N+1}}\right)^2, \label{action5}
\end{eqnarray}
and from Eqs. (\ref{trans}) and (\ref{dynpar}) the normalized
transition probability
\begin{eqnarray}
P(\delta x_1\rightarrow\delta x_N,N)=\Omega(N)^{-1}\times
\exp\left[-\frac{(a^2-1)(1-a^{-2N})}{2\Delta}\left(\frac{\delta
x_N-\delta x_1 a^{N-1}}{a^{N-1}-a^{-N+1}}\right)^2\right],
\end{eqnarray}
with dynamic partition function
\begin{eqnarray}
\Omega(N)=(a^{N-1}-a^{-N+1})\left(
\frac{2\pi\Delta}{(a^2-1)(1-a^{-2N})}\right)^{1/2}.
\end{eqnarray}
In the long time limit $N\rightarrow\infty$ the behavior of $S$,
$P$, and $\Omega$ depends on $a=f'(x_1^\ast)$. For $|a|<1$ we
obtain
\begin{eqnarray}
S\rightarrow\frac{1}{2}(1-a^2)\left(\frac{\delta x_N}{a}\right)^2,
\end{eqnarray}
and thus the stationary distribution
\begin{eqnarray}
P_{\text{st}}\propto\exp\left[-\frac{1}{2\Delta}(1-a^2)
\left(\frac{\delta x_N}{a}\right)^2\right].
\end{eqnarray}
%
%%%%%%%%%%%%%%%%%%%%%%%%%%%%%%%%%%%%%%%%%%%%%%%%
%%%%%%%%%%%%%%%%%%%%%%%%%%%%%%%%%%%%%%%%%%%%%%%%
\subsubsection{The noisy fixed point}
%%%%%%%%%%%%%%%%%%%%%%%%%%%%%%%%%%%%%%%%%%%%%%%%
%%%%%%%%%%%%%%%%%%%%%%%%%%%%%%%%%%%%%%%%%%%%%%%%
The noisy fixed point
$(x^\ast,\pi^\ast)=(x_2^\ast,x_2^\ast-f(x_2^\ast))$,
$f'(x_2^\ast)=1$ is elliptic for $|\pi_2^\ast f''(x_2^\ast)-2|<2$
and hyperbolic for $|\pi_2^\ast f''(x_2^\ast)-2|>2$. In the
hyperbolic case the discussion in the deterministic case applies,
we leave the details to the reader. In the elliptic case the
complex eigenvalues are determined by Eq. (\ref{eigen}). The orbits
are periodic and given by
\begin{eqnarray}
&&\delta x_n = A\cos(\omega n+\psi_1),
\\
&&\delta \pi_n = B\cos(\omega n+\psi_2),
\end{eqnarray}
where the frequency $\omega$ is given by
\begin{eqnarray}
\tan\omega=\frac{2-\pi_2^\ast f''_2}{ \sqrt{\pi_2^\ast
f''_2(4-\pi_2^\ast f''_2)}}.
\end{eqnarray}

In the hyperbolic case, the long time orbits near the fixed point
permits a derivation of the transition probability as in the case
of the deterministic fixed point on the $\pi=0$ manifold. In the
elliptic case, the finite time orbits also allow an evaluation of
$P$; we shall, however, not pursue this analysis here.

Summarizing, the 2D map describing the noisy 1D map within the
nonperturbative weak noise approach is characterized by at least
two fixed points. The noiseless deterministic fixed point is
hyperbolic. The character of the noisy fixed point(s)
depends on the magnitude of $\pi_2^\ast f''(x_2^\ast)$. In the
vicinity of the hyperbolic fixed point the orbits slow down and we
can derive a stationary probability distribution in the linear
regime. Near the elliptic fixed point the finite time orbits only
yield the short time character of the probability distributions.

Owing to the nonlinear character of the 2D map we anticipate that
the orbit structure near the hyperbolic fixed point breaks down
due to heteroclinic intersections and the formations of
foliations. In the vicinity of the elliptic fixed point we expect
to encounter the dissolution of the KAM tori into cantori and
chaotic seas with Arnold diffusion \cite{Ott93,Reichl87}. Some of
these features will be investigated in more detail in the next
section on the noisy logistic map.
%%%%%%%%%%%%%%%%%%%%%%%%%%%%%%%%%%%%%%%%%%%%%%
%%%%%%%%%%%%%%%%%%%%%%%%%%%%%%%%%%%%%%%%%%%%%%
\section{1D Noisy Maps}
%%%%%%%%%%%%%%%%%%%%%%%%%%%%%%%%%%%%%%%%%%%%%%
%%%%%%%%%%%%%%%%%%%%%%%%%%%%%%%%%%%%%%%%%%%%%%
In this section we embark on a discussion of specific noise-driven
1D maps which in the absence of noise exhibits nonlinear behavior
such as period doubling and transitions to chaotic behavior.
%%%%%%%%%%%%%%%%%%%%%%%%%%%%%%%%%%%%%%%%%%%%%%
%%%%%%%%%%%%%%%%%%%%%%%%%%%%%%%%%%%%%%%%%%%%%%
\subsection{The linear map}
%%%%%%%%%%%%%%%%%%%%%%%%%%%%%%%%%%%%%%%%%%%%%%
%%%%%%%%%%%%%%%%%%%%%%%%%%%%%%%%%%%%%%%%%%%%%%
It is instructive to briefly consider the linear map
\begin{eqnarray}
x_{n+1}=r x_n,
\end{eqnarray}
with control parameter $r$. The Jacobian $J=dx_{n+1}/dx_n=r$ and
for $r<1$ the map is length-shrinking with a stable fixed point at
$x^\ast=0$; for $r=0$ the map is length-preserving with a line of
fixed points. For $r>0$ the map is length-expanding, the fixed
point at $x^\star$ is unstable and the orbit recedes to infinity.
In the presence of noise we obtain the noisy map
\begin{eqnarray}
x_{n+1}=r x_n + \xi_n,
\end{eqnarray}
corresponding to the 2D area-preserving map
\begin{eqnarray}
&&x_{n+1}=r x_n +r^{-1}\pi_n,
\\
&&\pi_{n+1} = r^{-1}\pi_n.
\end{eqnarray}
This map possesses a hyperbolic fixed point at $(x^\ast,\pi^\ast)$
and the discussion in Sec. II applies.
%%%%%%%%%%%%%%%%%%%%%%%%%%%%%%%%%%%%%%%%%%%%%%
%%%%%%%%%%%%%%%%%%%%%%%%%%%%%%%%%%%%%%%%%%%%%%
\subsection{The logistic map}
%%%%%%%%%%%%%%%%%%%%%%%%%%%%%%%%%%%%%%%%%%%%%%
%%%%%%%%%%%%%%%%%%%%%%%%%%%%%%%%%%%%%%%%%%%%%%
We now turn to the well-known 1D logistic map
\begin{eqnarray}
x_{n+1} = r x_n(1-x_n),
\end{eqnarray}
which has fixed points at $x^\ast=0$ and $x^\ast=1-1/r$. For $r>3$
the fixed point bifurcates to period-2 fixed points and in the
regime $3<r<r_c$, $r_c\approx 3.57$ the map passes through a
period-doubling scenario exhibiting universal scaling behavior
\cite{Feigenbaum78,Feigenbaum79}.

In the presence of noise we have the stochastic map
\begin{eqnarray}
x_{n+1} = r x_n(1-x_n) + \xi_n,
\end{eqnarray}
and the associated area preserving 2D map
\begin{eqnarray}
&&x_{n+1} = r x_n(1-x_n) + \frac{\pi_n}{r(1-2x_n)},
\\
&&\pi_{n+1}=\frac{\pi_n}{r(1-2x_n)}.
\end{eqnarray}
Since $f(x)=rx(1-x)$, $f'(x)=r(1-2x)$, and $f''(x)=-2r$ and
according to the general discussion in Sec. II the 2D map has two
deterministic fixed points on the $x$- axis and a noisy fixed
point in the lower half $x-\pi$ plane:
\begin{eqnarray}
&&(x^\ast,\pi^\ast)=(0,0),~~~~~~~~~~~~~~~~~~~~~~~~~~~~~~~~~\text{(FP1)}
\\
&&(x^\ast,\pi^\ast)=(1-1/r,0),~~~~~~~~~~~~~~~~~~~~~~~~~\text{(FP2)}
\\
&&(x^\ast,\pi^\ast)=((r-1)/2r,-(1-r)^2/4r).~~~~~\text{(FP3)}
\end{eqnarray}
The trace and eigenvalues characterizing the stability and
properties of the fixed points are given by Eqs. (\ref{trace}) and
(\ref{eival}); likewise, the invariant manifolds follow from
$(\delta x,\delta\pi)\propto(1,-(f'^2-\pi f''-f'\lambda_\pm))$.
The fixed point form a triad. The traces associated with the fixed
points are given by
\begin{eqnarray}
&&t_1= r+\frac{1}{r},
\\
&&t_2=2-r+\frac{1}{2-r},
\\
&&t_3=2-\frac{(1-r)^2}{2}.
\end{eqnarray}

We can now read off the character of the fixed points and
establish the correspondence with the bifurcations of the
noiseless 1D map. First, we infer that for all $r>0$ (we only
consider positive $r$) the deterministic fixed points FP1 and FP2
on the $x$-axis are hyperbolic, whereas the noisy fixed point is
elliptic for $r<r_t$ and hyperbolic for $r>r_t$, where
$r_t=1+\sqrt{8}\approx 3.8284$ corresponds to the period 3 tangent
bifurcation for the 1D map.

Let us first concentrate at the first period doubling point at
$r=3$. The Jacobian matrix (\ref{tmap}) for the 2D map of the
logistic equation is of the form
\begin{eqnarray}
M=\left(
\begin{array}{cc}
r-2rx^\ast + \frac{2\pi^\ast}{r(1-2x^\ast)^2}
& \frac{1} {r-2rx^\ast}\\
\frac{2\pi^\ast}{r(1-2x^\ast)^2}&\frac{1}{r-2rx^\ast}
\end{array}
\right).
\end{eqnarray}
For the fixed point FP2 (where $\pi^\ast=0$) at $r=3$ it then
takes the form
\begin{eqnarray}
M=\left(
\begin{array}{cc}
-1
& -1\\
0&-1
\end{array}
\right).
\end{eqnarray}
\begin{figure}
\includegraphics[width=0.9\hsize]
%{c:/user/manus/figs/162-223.eps}
{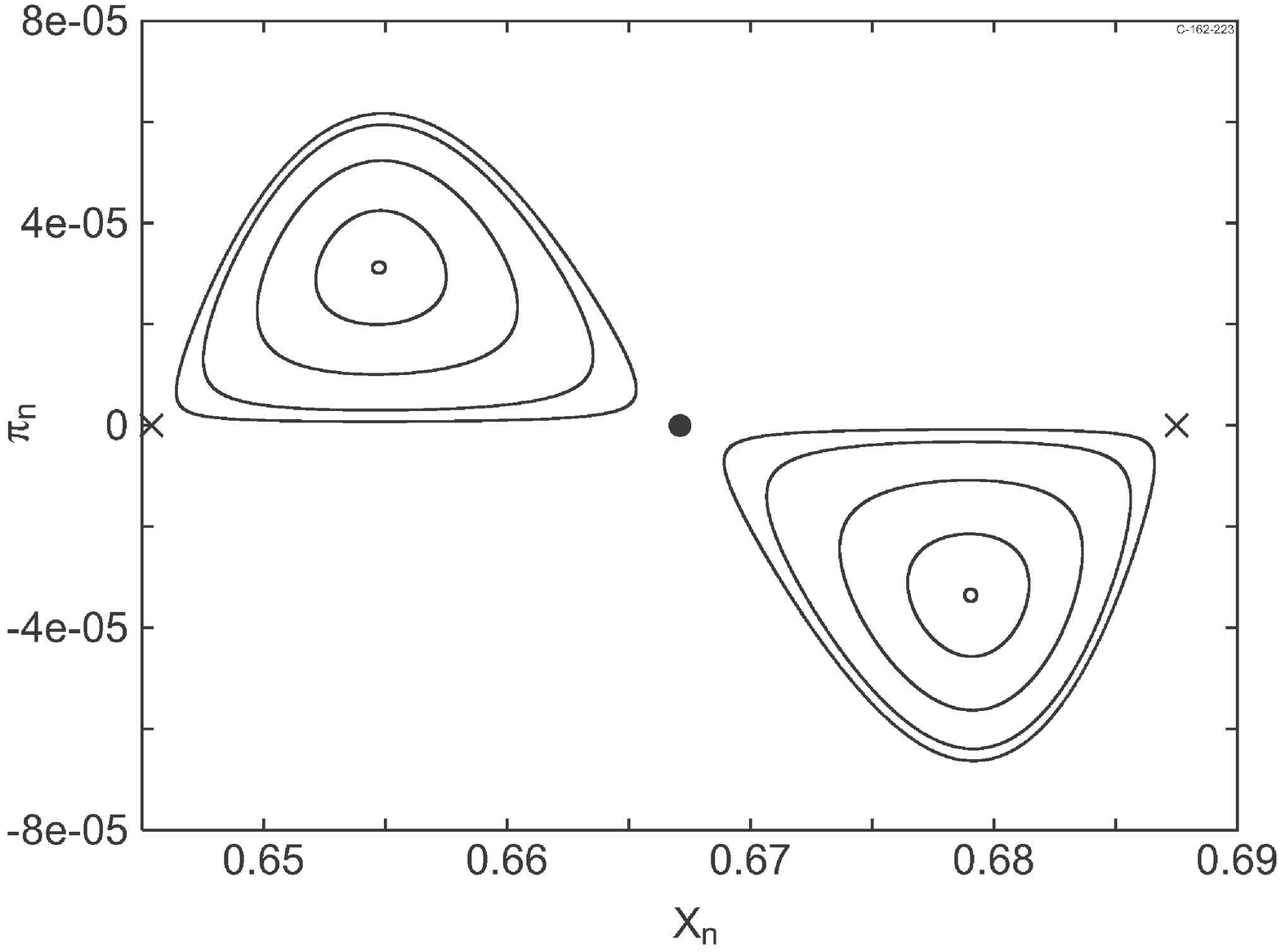} \caption{The phase plane $(x_n, \pi_n)$ for
$r=3.004$ just above the period-doubling into the two-cycle. The
black dot is the now (hyperbolic) unstable fixed point; the
crosses are the two-cycle on the x-axis and the KAM curves
encircle the elliptic two-cycle.} \label{fig5}
\end{figure}
This corresponds to a degenerate node where the two
eigendirections coincide along the x-axis. For $r<3$ the
eigenvalues are $\lambda_1 <-1$ and $\lambda_2 > -1$,
respectively. At $r=3$ the eigenvalues collide in -1 signalling
that two period-doublings are about to take place. In
Fig.~\ref{fig5} we show this behavior just above $r=3$, at
$r=3.004$. The black dot represents the (unstable) fixed point
FP2, which seen in the 2D plane now is a hyperbolic fixed point.
The crosses determines the ``usual" stable period two cycle.
Simultaneously, a period two cycle of elliptic points has been
spinning off. Initially, along the eigendirection in the x-axis
but quickly moving nonlinearly into the plane. We have numerically
followed the upper elliptic two-cycle point as far as we could for
increasing $r$. The KAM surfaces around it gradually dissolves and
to the best of our knowledge without the cycle point undergoes
further period-doubling's. Fig.~\ref{fig6} shows this two-cycle
point at $r=3.608$, with only a few remaining KAM surfaces around
it, seen on a tiny scale. %%
\begin{figure}
\includegraphics[width=0.9\hsize]
%{c:/user/manus/figs/162-226.eps}
{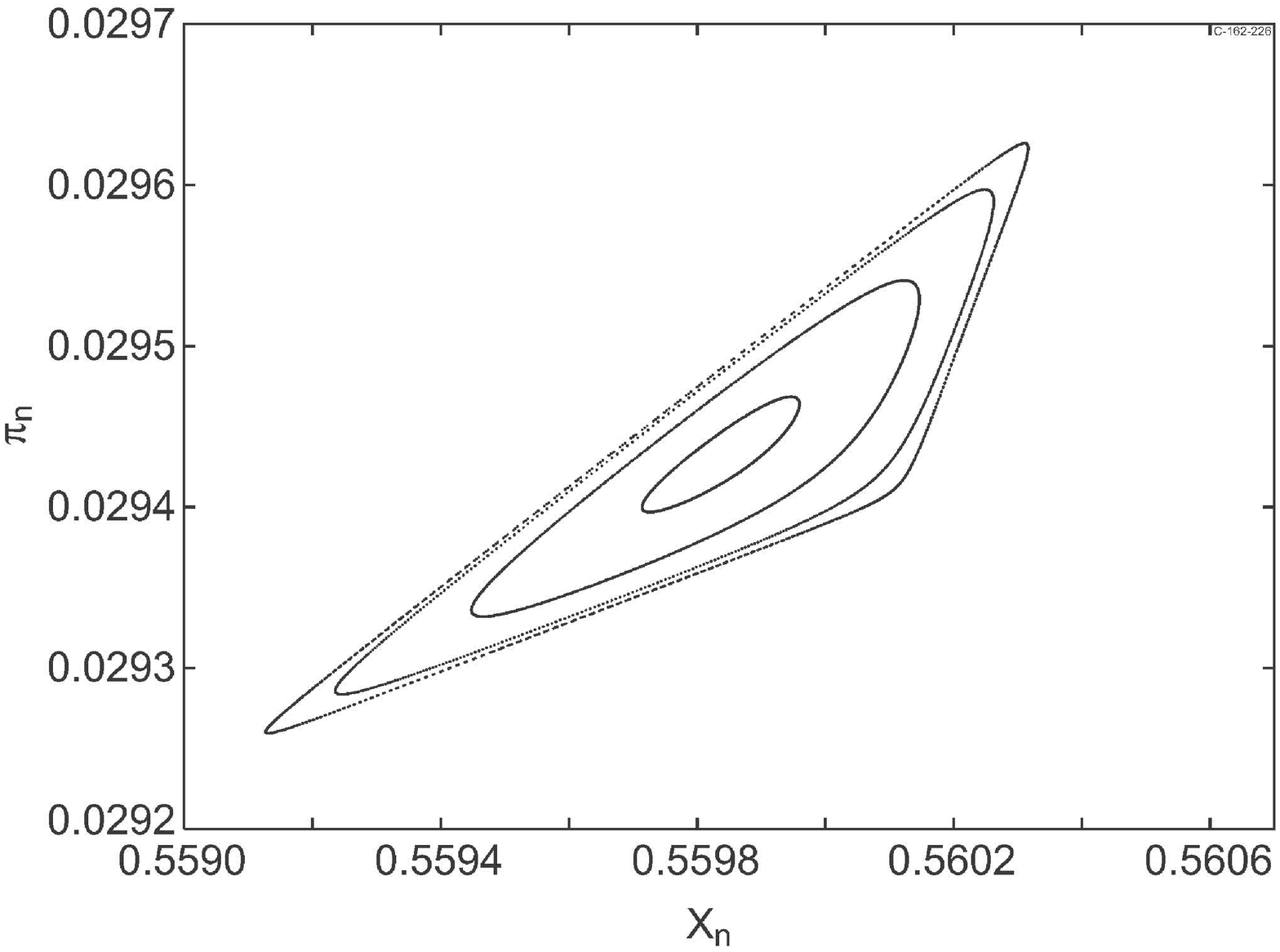} \caption{One of the elliptic two-cycle points
for $r=3.6084$ in the ``chaotic sea" of the energy surface. The
KAM surfaces remain only on very small scales close to the cycle
point.} \label{fig6}
\end{figure}
\begin{figure}
\includegraphics[width=0.9\hsize]
%{c:/user/manus/figs/162-225.eps}
{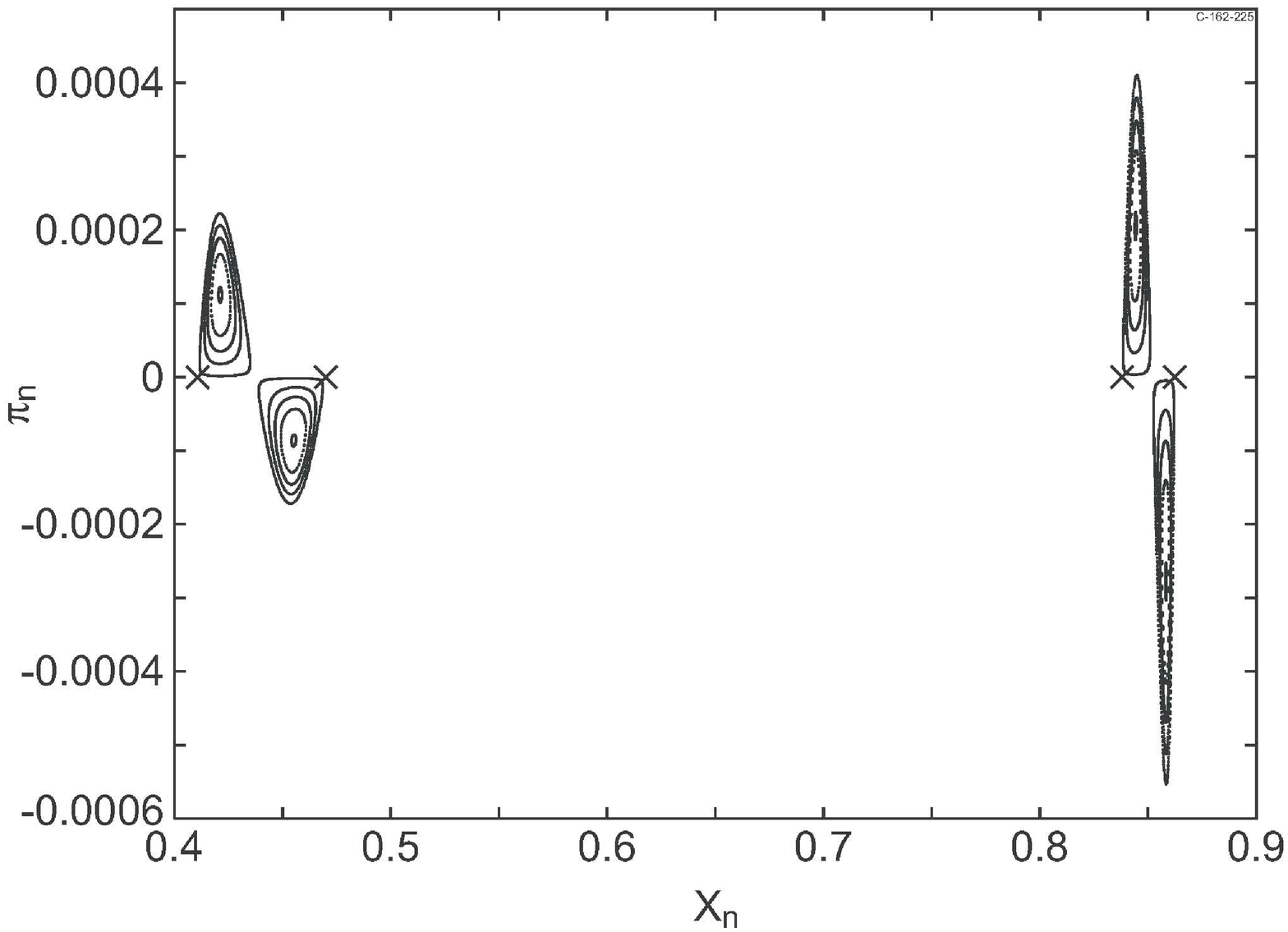} \caption{The behavior at $r=3.462$ just above
the period-doubling into the four-cycle. Note, the usual stable
four-cycle on the axis and the associated elliptic four-cycle.}
\label{fig7}
\end{figure}
Furthermore, for increasing values of $r$, what happens is a
continuous spinning off of elliptic cycles in close association
with the standard period-doubling sequence on the x-axis. We show
this in Fig.~\ref{fig7} (for the parameter value $r=3.462$) for
the period-doubling to the four-cycle occurring at $r \approx
3.449$. Note again the normal stable four-cycle, represented by
crosses, on the x-axis and the associated four-cycle of elliptic
points in the plane. To substantiate this picture even further, we
finally study the behavior just above the period-doubling into the
stable eight-cycle. At $r=3.55$, this is shown in Fig.~\ref{fig8}
with the same structure of the stable eight-cycle on the axis with
the elliptic eight-cycle in the plane. We thus conclude, that this
process will continue in the same fashion, with elliptic $2^n$
cycle moving away from the x-axis into the plane at each
bifurcation point. %%

\begin{figure}
\includegraphics[width=0.9\hsize]
%{c:/user/manus/figs/162-224.eps}
{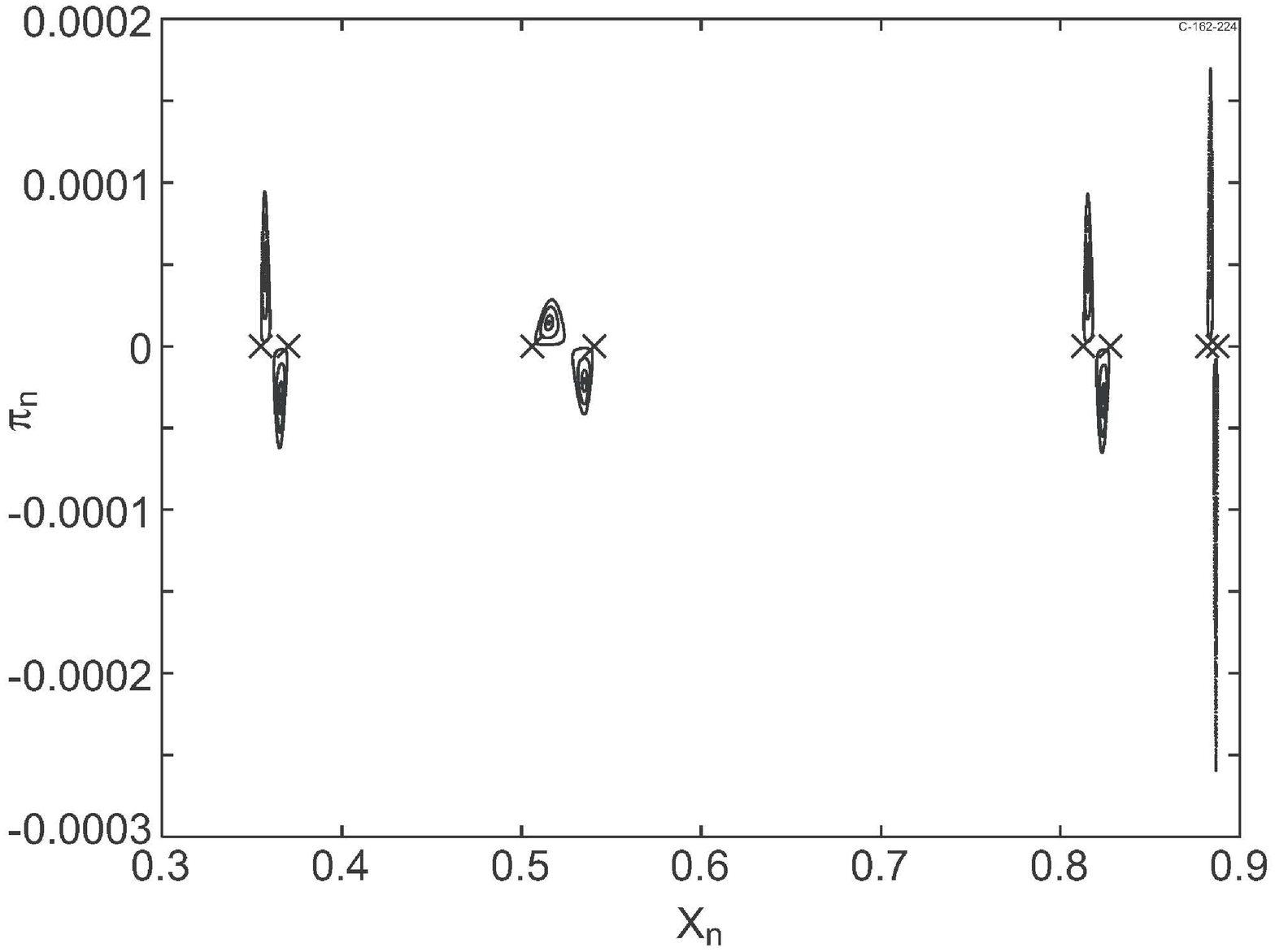} \caption{ The phase plane at $r=3.55$ above
the period-doubling into the eight-cycle, where the crosses
indicate the stable eight-cycle.} \label{fig8}
\end{figure}
For $r>r_c $ the 1D map approaches a strange attractor with
positive Liapunov exponent. In the 2D map this is reflected by
orbits approaching the x-axis subject to infinitely many
foliations in order to preserve the area.
%%%%%%%%%%%%%%%%%%%%%%%%%%%%%%%%%%%%%%%%%%%%%%
%%%%%%%%%%%%%%%%%%%%%%%%%%%%%%%%%%%%%%%%%%%%%%
\section{Summary and conclusion}
%%%%%%%%%%%%%%%%%%%%%%%%%%%%%%%%%%%%%%%%%%%%%%
%%%%%%%%%%%%%%%%%%%%%%%%%%%%%%%%%%%%%%%%%%%%%%
We have attacked the old problem of noise in 1D maps, in
particular the logistic equation, in a new way. Starting out with
a discrete version of the standard Langevin equation, we apply a
nonperturbative weak noise approach to map the stochastic equation
onto a deterministic 2D map. This map is area-preserving and the
added dimension plays the role of the noise field. We have derived
general properties of this 2D map and applied it to two well know
systems, the trivial linear map and the logistic map. In both
cases we find, that in addition to the standard fixed points in
the usual variable, there are additional fixed points and higher
order cycles out in the plane of finite noise amplitude. These
fixed and cycle points are either hyperbolic or elliptic. In the
case of the logistic map we find that each standard
period-doubling on the x-axis is always associated with a
period-doubling out in the plane. These cycles in the plane of
order $2^n$ are born as elliptic points moving away from the cycle
points on the x-axis at the bifurcation points.
%%%%%%%%%%%%%%%%%%%%%%%%%%%%%%%%%%
\begin{acknowledgments}
The authors wish to thank Niels S{\o}ndergaard for fruitful
discussions. The present work has been supported by the Danish
Research Council.
\newpage
\newpage
\end{acknowledgments}

%\bibliography{c:/user/manus/bib/articles,c:/user/manus/bib/books}
\end{document}